# Exploring proteomic signatures in sepsis and non-infectious systemic inflammatory response syndrome

Adolfo Ruiz-Sanmartín[1,2,3], Vicent Ribas[4], David Suñol[4], Luis Chiscano-Camón[1,2,3], Laura Martín[2,3], Iván Bajaña[2,3], Juliana Bastida[2,3], Nieves Larrosa[3,5,7], Juan José González[3,5,7], M Dolores Carrasco[9], Núria Canela[4], Ricard Ferrer[1,2,3], Juan Carlos Ruiz-Rodríguez[1,2,3]

1. Departament de Medicina, Universitat Autònoma de Barcelona, Barcelona, Spain.
2. Intensive Care Department, Vall d'Hebron University Hospital, Vall d'Hebron Barcelona Hospital Campus, Barcelona, Spain.
3. Shock, Organ Dysfunction and Resuscitation (SODIR) Research Group, Vall d'Hebron Research Institute, Barcelona, Spain.
4. Eurecat, Centre Tecnològic de Catalunya, Digital Health Unit, Barcelona, Spain.
5. Department of Clinical Microbiology, Vall d'Hebron University Hospital, Vall d'Hebron Barcelona Hospital Campus, Barcelona, Spain.
6. Eurecat, Centre Tecnològic de Catalunya, Centre for Omic Sciences (COS), Joint Unit URV-EURECAT, Unique Scientific and Technical Infrastructures (ICTS), Reus, Spain.
7. Department of Genetics and Microbiology, Universitat Autònoma de Barcelona, Barcelona, Spain.
8. CIBERINFEC, ISCIII – CIBER de Enfermedades Infecciosas, Instituto de Salud Carlos III, Madrid, Spain.
9. Post-cardiac Surgery Unit. Department of Intensive Care, Vall d'Hebron University Hospital, Vall d'Hebron Barcelona Hospital Campus, Barcelona, Spain.

Corresponding autor: Adolfo Ruiz-Sanmartín 1338842@uab.cat

**ABSTRACT**

**Background:** The search for new biomarkers that allow an early diagnosis in sepsis has become a necessity in medicine. The objective of this study is to identify potential protein biomarkers of differential expression between sepsis and non-infectious systemic inflammatory response syndrome (NISIRS).

**Methods:** Prospective observational study of a cohort of septic patients activated by the Sepsis Code and patients admitted with NISIRS, during the period 2016-2017. A mass spectrometry-based approach was used to analyze the plasma proteins in the enrolled subjects. Subsequently, using recursive feature elimination (RFE) classification and cross-validation with a vector classifier, an association of these proteins in patients with sepsis compared to patients with NISIRS. The protein-protein interaction network was analyzed with String software.

**Results:** A total of 277 patients (141 with sepsis and 136 with NISIRS) were included. After performing RFE, 25 proteins in the study patient cohort showed statistical significance, with an accuracy of 0.960 (95% CI: 0.936–0.983), specificity of 0.920 (95% CI: 0.859–0.980), sensitivity of 0.973 (95% CI: 0.945–1.00), and an AUC of 0.985 (95% CI: 0.972–0.997). Of these, 14 proteins (vWF, PPBP, C5, C1RL, FCN3, SAA2, ORM1, ITIH3, GSN, C1QA, CA1, CFB, C3, LBP) have a greater relationship with sepsis while 11 proteins (FN1, IGFALS, SERPINA4, APOE, APOH, C6, SERPINA3, AHSG, LUM, ITIH2, SAA1) are more expressed in NISIRS.

**Conclusion:** There are proteomic patterns associated with sepsis compared to NISIRS with different strength of association. Advances in understanding these protein changes may allow for the identification of new biomarkers or therapeutic targets in the future.

**Key words:** Sepsis, Septic shock, SIRS, Proteomics, Omics, Diagnosis

# INTRODUCTION

Sepsis is known as a clinical syndrome where life-threatening organ dysfunction occurs due to a dysregulated host response to infection. The severity of sepsis varies significantly with the response and degree of organ dysfunction. Severe cases of sepsis, during which hypotension persists even after adequate fluid resuscitation and lactate levels > 2 mmol/L, are classified as septic shock [1]. Despite advances in diagnosis and treatment, sepsis remains one of the leading causes of morbidity and mortality worldwide, with a mortality rate ranging around 30-50% [2,3].

Current decisions regarding sepsis diagnosis and treatment are primarily based on Sequential Organ Failure Assessment (SOFA) and Quick SOFA (qSOFA), but their sensitivity and accuracy are known to be lacking [4,5]. C-reactive protein (CRP), procalcitonin (PCT), interleukin-6 (IL-6), and other biomarkers are also used for sepsis detection. Most of these biomarkers can reflect the immune system's state and stages of the inflammatory cascade, being protein molecules with negatively regulated gene expression. CRP is frequently used to identify infections and sepsis. However, CRP cannot accurately reflect the severity of infection and sepsis because it increases during a minor infection or remains elevated even after the temporal course of the infection. Additionally, CRP levels can also rise during an inflammatory response to non-infectious events, trauma, tumorigenesis, or surgical interventions. These findings suggested that CRP lacks specificity as an early-stage sepsis biomarker [6,7]. PCT is likely the best-suited biomarker for infection at present, and it has even been proposed as a prognostic factor for sepsis progression [8] and a guide for antibiotic treatment duration [9]. However, it is hindered by false positives in non-infectious inflammation settings and a rather delayed induction (4 to 12 hours with a half-life of 22 to 35 hours) during the host's response to infection [10,11]. Other biomarkers such as presepsin or pro-ADM have also been proposed as promising biomarkers in sepsis [12-14].

A deep understanding of the molecular and cellular mechanisms involved in sepsis is essential for more accurate and early diagnosis, as well as the development of new therapeutic strategies [15]. In this context, proteomics (a discipline of molecular biology that studies the complete set of proteins expressed in a cell, tissue or organ) has emerged as a powerful and promising tool in the study of complex protein interactions underlying sepsis [16]. The use of techniques like two-dimensional electrophoresis, liquid chromatography, and mass spectrometry have led to the identification of specific biomarkers for early diagnosis and prognosis of sepsis. These biomarkers can assist clinicians in swiftly identifying high-risk patients and making more precise therapeutic decisions. The main goal of proteomics in the study of sepsis is to identify specific biomarkers and key molecular pathways involved in disease progression and prognosis. The identification of accurate and sensitive biomarkers would enable early diagnosis and more effective monitoring of sepsis, potentially improving clinical outcomes and reducing associated mortality rates [17-19].

The hypothesis of this study is that there are proteomic patterns in patients with sepsis that differentiate them from patients with NISIRS.

The objective of this study is to identify potential protein biomarkers of differential expression between sepsis and NISIRS.

**METHOD**

**Study design and ethical approval**

This is a prospective, observational, single-center study with two study populations. One study group with septic patients who met the criteria for activation of the Vall d'Hebron University Hospital in-hospital Sepsis code [20] (ISC) between April 2016 and January 2018. The second study group included patients admitted to the Intensive Care Unit who met criteria for Systemic Inflammatory Response Syndrome (SIRS) without evidence of infection [21]. The study was approved by the Clinical Research Ethics Committee of Vall d'Hebron University Hospital [PR (AG) 11-2016, PR (AG) 336-2016, PR (AG) 210/2017], and written informed consent was obtained from all participants. The study fully adhered to the General Data Protection Regulation (Regulation (EU) 2016/679) and was conducted in accordance with the ethical standards outlined in the 1964 Declaration of Helsinki and its subsequent amendments.

**Inclusion and exclusion criteria**

The inclusion criteria for patients with NISIRS were adult patients ≥ 18 years who presented with two or more of the following variables: (1) White blood cell count >12,000/mm3 or <4,000/mm3, or >10% immature cells, (2) the presence of hyperthermia (axillary temperature >38.3 ºC) or hypothermia (axillary temperature <36.0ºC), and/or tachycardia (>100 beats per minute), tachypnea (>30 breaths per minute), or desaturation (SpO2 <90%) and (3) absence of infection. The inclusion criteria for the septic patients group encompassed adult patients ≥ 18 years of age with suspected or documented infection and the presence of, at least, one of the following sets of variables, as outlined by the ISC: (1) an acute alteration in the level of consciousness not explained by other clinical conditions, or (2) the presence of hyperthermia (axillary temperature >38.3 ºC) or hypothermia (axillary temperature <36 ºC), and/or tachycardia (>110 beats per minute), tachypnea (>30 breaths per minute) or desaturation (SpO2 <90%), as well as arterial hypotension (systolic blood pressure <90 mmHg or mean arterial pressure <65 mmHg or >40 mmHg decreased in baseline systolic blood pressure).

Exclusion criteria include non-adult patients, pregnant women or patients from whom a blood sample or written informed consent could not be obtained.

**Data collection and biomarker measurements**

Following patient enrollment in the study, demographic data were recorded, and a venous or arterial blood sample was obtained at the time of the initial visit for routine laboratory value assessments. Additionally, samples were collected for microbiological cultures in patients suspected of having sepsis. Clinical scores (such as SOFA) were retrospectively calculated whenever feasible at the time of enrollment. Measurements of CRP using an immunoturbidimetric test and lactate using an enzymatic color test were performed on these samples. The collected samples were frozen at -80ºC and stored in a Sepsis Bank of Vall d'Hebron University Hospital Biobank with appropriate ethics approval for subsequent analysis in accordance with clinical laboratory protocols.

The proteomic analysis of the samples by mass spectrometry is provided in supplementary material 1.

**STATISTICAL ANALYSIS**

Demographic, clinical, and laboratory data were reported as mean ± standard deviation or median with interquartile range as appropriate, and categorical variables as numbers and percentages. The Student's t-test was used for parametric quantitative variables, Mann-Whitney U test for non-parametric quantitative variables, and Chi-square test for qualitative variables. Statistical significance was determined at $p < 0.05$. The statistical analysis was performed using SPSS 18.0 software (SPSS Inc., Chicago, IL, USA).

In the proteomic study, prior to conducting any statistical analysis, each protein was standardized, and missing values were imputed using the k-nearest neighbor (KNN) method for proteins with less than 25% missing values. Proteins with major missing assignments were excluded from the study. The Kruskal-Wallis method with Benjamini-Hochberg false discovery rate (FDR) correction test ($p < 0.05$) was used to assess differences between distributions. The Benjamini-Hochberg procedure was applied to control the FDR. Statistical analyses were conducted in Python 3.8 using the pandas, sklearn, spicy, and stats models libraries.

**Protein selection**

Protein selection was carried out in five steps. In the first step data availability was assessed for completeness and consistency. In this step, data with missingness percentage greater than 25% has been censored. This process filtered out 65 out of 177 proteins. After that, two different datasets have been generated: one where missing values have been imputed with the k-nearest neighbors (KNN) method and a second one with no imputation.

The second step consisted in a statistical analysis with the Kruskal-Wallis method with Benjamini-Hochberg false discovery rate (FDR) correction over the two datasets. The statistical analysis yielded the same list of 78 proteins with statistically significant expression values between Sepsis and NISIRS.

The third step consists of a recursive feature elimination (RFE) with a logistic regression over the two datasets generated in the first step outlined above. RFE was applied with a 10-fold cross validation approach with stratified 80-20% splits for training and validation. The resulting predictions in validation between the two datasets were assessed with the MacNemar statistical test with a p-value of 0.3. Since there are no statistically significant differences between the results for the two datasets, it was decided to continue the experimental setup with the imputed dataset.

The fourth step consisted in assessing the discriminative power of the protein list obtained in the previous step. This step has been implemented with a logistic regression with a 10-fold cross validation approach with stratified 80-20% splits where accuracy, sensitivity, specificity, and AUC have been reported with 95% confidence intervals (95% CI). The logistic regression coefficients have also been reported with 95% confidence intervals, z-score, and p-values.

In the fifth and final step, the coefficients of the logistic regression were analyzed using their additive Shapley explanations (SHAP values in summary). Proteins with positive Shapley values were associated with sepsis, while negative Shapley values were associated with NISIRS. The strength of association between the Shapley value and the outcome (sepsis and NISIRS) was measured by the magnitude of these Shapley values.

Protein selection was performed in Python 3.8 using the standard libraries pandas and scikit-learn. The protein-protein interaction network was analyzed using String v 11.0b software (https://string-db.org/).

**RESULTS**

**Characteristics of the study population**

A total of 277 patients were included in this study, with 141 patients in the sepsis group and 136 in the NISIRS group. The demographic and clinical data of the patients are shown in Table 1. For the sepsis group, the most common infection focus was urinary 49 (34.8%), followed by respiratory 47 (33.3%), and abdominal 44 (31.2%). In the NISIRS group, 107 (78.67%) patients had been admitted post-cardiac surgery, 13 (9.55%) were lung transplant recipients, 5 (3.67%) were liver transplant recipients, 4 (2.95%) had hemorrhagic shock, 3 (2.20%) were kidney transplant recipients, 2 (1.47%) were polytrauma patients, 1 (0.75%) had splenic hematoma, and 1 (0.75%) patient had acute pancreatitis.

**Table 1**. Characteristics of the study population.

| Characteristics | Total (n=277) | Sepsis (n=141) | NISIRS (n=136) | P |
|---|---|---|---|---|
| **Male**, n (%) | 162 (58.48) | 85 (60.28) | 77 (56.61) | 0.53 |
| **Age** years (m±SD) | 63.38±15.61 | 63.9±15.6 | 62.8±15.5 | 0.54 |
| **SOFA score** median (25th,75th) | 5 (3,7) | 7 (5,8) | 3 (2,6) | <0.05 |
| **Norepinephrine**, n(%) | 121 (43.68) | 76 (53.90) | 45 (33.08) | <0.05 |
| **ICU admission**, n (%) | 206 (74.4) | 70 (49.64) | 136 (100) | < 0.05 |
| **Mechanical Ventilation**, n (%) | 177 (63.9) | 41 (29.07) | 136 (100) | <0.05 |
| **Leucocytes x $10^6$**, (mean ± SD) | 14118±9149 | 13501±11021 | 14757±661 | 0.25 |
| **Platelets x $10^9$**, median (25th,75th) | 130.85 (116.0-227.5) | 184.00(114.0-278.5) | 157.00(119.5-195.2) | <0.05 |
| **Lactate mmol/L**, median (25th,75th) | 1.9 (1.4,3.1) | 2.5 (1.8,4.1) | 1.5 (1.0,1.9) | <0.05 |
| **CRP mg/dL**, (mean ± SD) | 14.61±11.90 | 21.77±12.58 | 7.58±5.06 | <0.05 |
| **Mortality**, n(%) | 35 (12.6) | 33 (24.2) | 2 (1.4) | <0.05 |

CRP: C-reactive protein.

**Proteomic study results**

Initially, a total of 110 proteins were identified by mass spectrometry for differential proteomic evaluation between NISIRS and Sepsis. Among them, 25 proteins in the study patient cohort showed statistical significance, with an accuracy of 0.960 (95% CI: 0.936 – 0.983), specificity of 0.920 (95% CI: 0.859 – 0.980), sensitivity of 0.973 (95% CI: 0.945 – 1.00), and an AUC of 0.985 (95% CI: 0.972 – 0.997). The analyzed proteins are presented in Table 2.

**Table 2.** Differentiated proteins analyzed between sepsis and NISIRS

| **Proteins studied in sepsis and NISIRS patients** | |
|---|---|
| **C1RL** - Complement C1r subcomponent-like protein | **PPBP** - Connective tissue-activating peptide III |
| **C3** - Complement C3c alpha' chain fragment 1 | **VWF** - Von Willebrand antigen 2 |
| **C5** - Complement C5 alpha' chain | **AHSG** - Alpha-2-HS-glycoprotein chain A |
| **C6** - Complement component C6 | **FN1** - Fibronectin |
| **CFB** – Complement factor B Ba fragment | **CA1** - Carbonic anhydrase 1 |
| **APOE** - Apolipoprotein E | **LUM** - Lumican |
| **APOH** – Beta-2-glycoprotein 1 | **SAA1** - Serum amyloid protein A |
| **FCN3** - Ficolin-3 | **SAA2** – Serum amylod A-2 protein |
| **GSN** - Gelsolin | **ORM1** - Alpha-1-acid glycoprotein 1 |
| **SERPINA3** - Alpha-1-antichymotrypsin | **IGFALS** – Insulin-like growth factor-binding protein complex acid labile subunit |
| **SERPINA4** - Kallistatin | |
| **LBP** - Lipopolysaccharide-binding protein | **C1QA** – Complement C1q subcomponent |

| **ITIH2** – Inter-alpha-trypsin inhibitor heavy chain H2 | subunit A |
| **ITIH3** - Inter-alpha-trypsin inhibitor heavy chain H3 | |

Of the 25 proteins found in this study, ten are involved in the regulation of proteolysis (SERPINA4, ITIH2, ITIH3, SERPINA3, FN1, APOE, C3, C5, GSN and AHSG), nine in innate immune response (LBP, FCN3, C3, C5, C6, GSN, CFB and C1RL), seven in complement activation (C1RL, C3, C5, C6, FCN3, C1QA and CFB), two in response to lipopolysaccharides (LBP and PPBP), four in blood coagulation (APOH, SAA1, FN1, VWF), two in lipid metabolism (APOE, APOH), and six proteins serve other functions (PPBP, ORM1, IGFALS, CA1, SAA2 and LUM). The relationship among these proteins is presented in Fig 1.

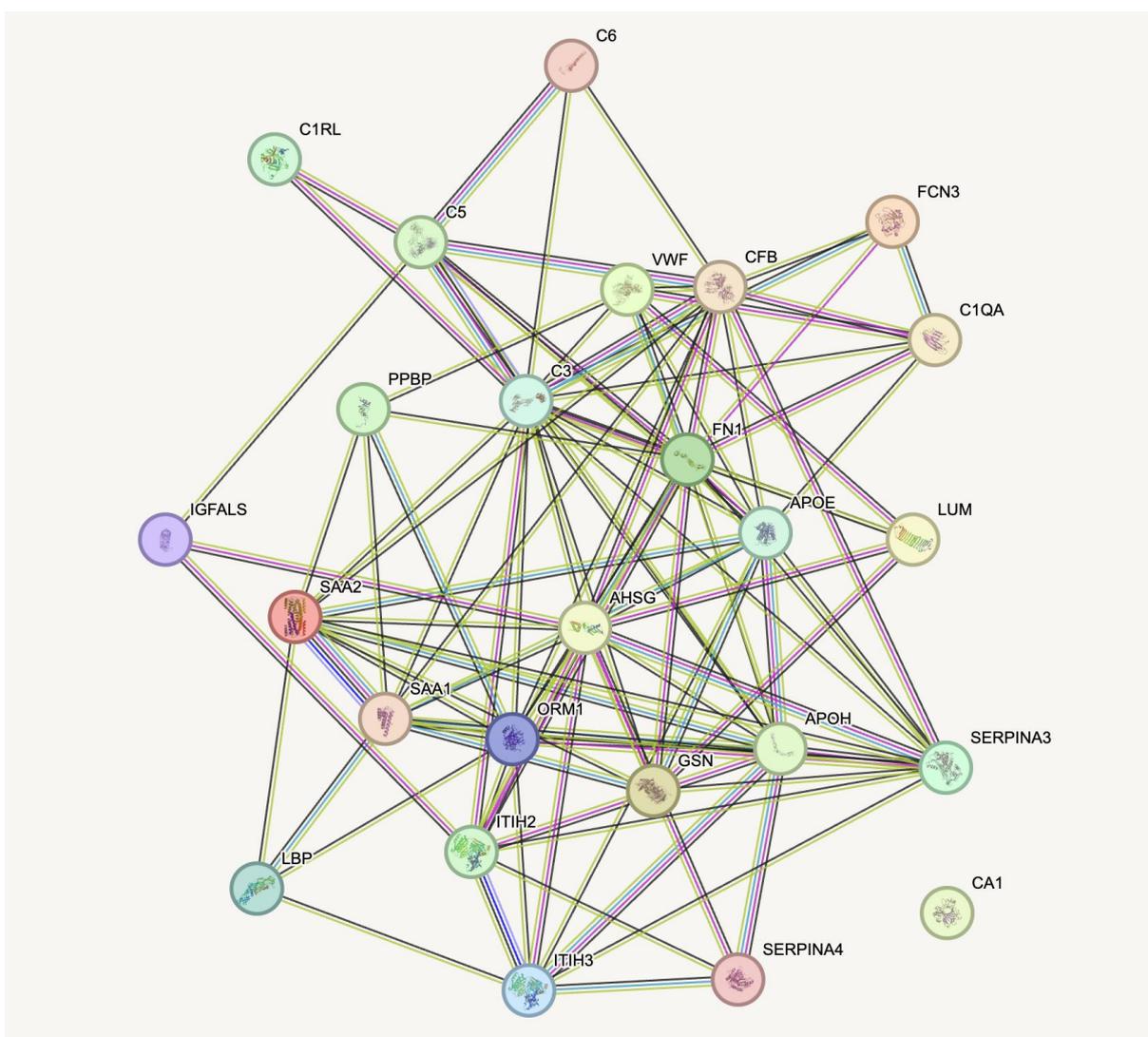

**Fig 1.** Network of physiological interactions between the different proteins analyzed (strings) (https://string-db.org/).

When applying the logistic regression model and analyzing its additive shape values, it was observed that the presence of 7 proteins with a higher association strength in the group of patients analyzed (Sepsis and NISIRS patients): PPBP (0.96), VWF (0.77), FN1 (0.57), CA1 (0.46), SERPINA4 (0.44), SAA2 ((0.44) and IGFALS (0.42). In contrast, proteins that show lower differential association strength between septic patients and patients with NSIRS are SAA1 (0.14), C6 (0.14), C3 (0.09) and ITIH2 (0.09) (Fig2).

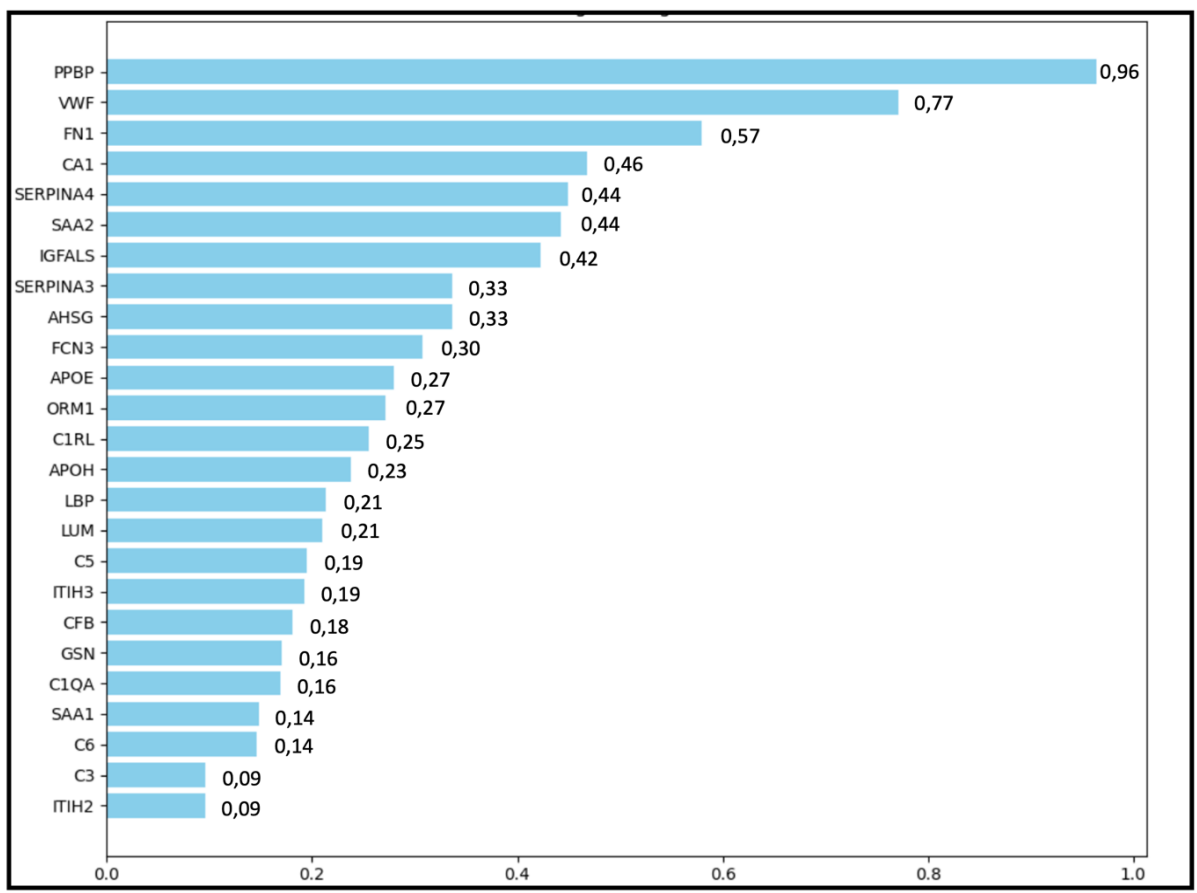

**Fig 2.** SHAP values graphic for logistic regression coefficients assessing the association strength of each protein in the overall set of patients studied.

Finally, after performing the analysis of the logistic regression coefficients using their additive Shapley explanations, we observed that the main proteins with the greatest association with sepsis are vWF (+1.32), PPBP (+1.31), C5 (+1.08), C1RL (+1.07), SAA2 (0.65), ORM1 (+0.64) and ITIH3 (0.64). The proteins that presented the highest negative value and, therefore, present the greatest association with NISIRS are FN1 (-1.30), IGFALS (-1.08), SERPINA4 (-1.04), APOE (-0.88), APOH (-0.82) and C6 (-0.80) (Fig3).

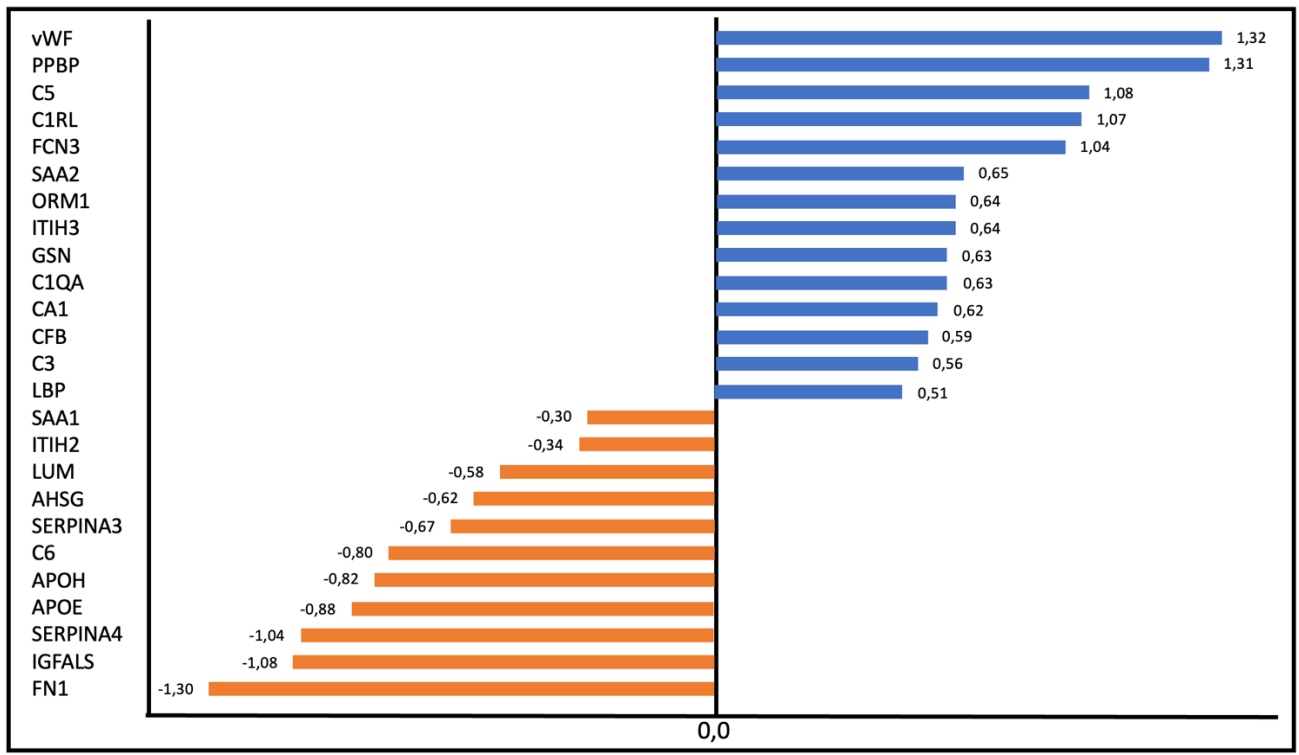

Fig 3. Coefficient values indicating tendency towards Sepsis (Blue) or NISIRS (Orange)

**DISCUSSION**

In this study, using proteomics techniques, we identified specific proteomic patterns associated with sepsis by comparing them with NISIRS patients. Thirty proteins were found with varying degrees of association with sepsis and NISIRS. The most relevant proteins associated to sepsis are vWF (1,32), PPBP (1,31), C5 (1,08), C1RL (1,07) and FCN3 (1,04). In contrast, the proteins found to have a greater relationship with NIRS are FN1 (-1,30), IGFALS (-1,08), SERPINA4 (-1,04) and APOE (-0,88). The identified proteins are involved in various physiological metabolic pathways, highlighting the complexity of sepsis as a process involving multiple pathways and biological systems.

In recent decades, multiple studies have observed different proteomic patterns in patients with sepsis, although few have compared them with groups having NISIRS. Shen et al. examined the plasma of 25 patients with SIRS and 25 with sepsis and found that seven proteins showed an increase in the plasma of patients with sepsis compared to SIRS, while three showed a decrease. The upregulated proteins included C4, CRP, plasminogen precursor, apolipoprotein A-II, plasma protease inhibitor C1 precursor, transthyretin precursor, and serum amyloid component P precursor. It was found that the APOA1 precursor, antithrombin-III precursor, and serum amyloid A-4 protein precursor were downregulated. This study revealed that complement and coagulation cascades were the most relevant pathways found to be altered in these samples [22]. None of these proteins were observed in our study. Su et al. conducted an analysis of urine from 15 patients with sepsis and 15 patients with SIRS. They identified a total of 130 proteins, of which 34 showed differential expression (with an increase greater than 1.5 times) and were associated with processes of inflammation, immunity, and structural or cytoskeletal aspects [23]. The authors, like in our study, find that proteins such as CA1 and C3 play a role in sepsis. Furthermore, we have observed that among these two proteins, CA1 exhibits a relatively stronger association with sepsis (0.62 vs. 0.56, respectively). More than twenty years ago, Stöve et al noted that the levels of C3a and the C3a/C3 ratio in the first 24 hours after the onset of sepsis were significantly higher in septic patients than in those with SIRS, while the levels of C3 in both groups were lower than in healthy donors. Additionally, the levels of C3a in septic patients decreased with appropriate treatment, and complement activation was much lower in patients with SIRS than in those with sepsis [24].

Most published studies on proteomics in sepsis have compared patients with healthy controls. For example, Chen Q et al. conducted a proteomic characterization in these groups and observed the expression of various proteins involved in different metabolic pathways, such as the acute inflammatory response, platelet degranulation, and the activation of myeloid cells in the immune response, among others. Among the analyzed proteins, they identified increased expression of LBP, S10A8, ORM1, FIBB, CRP, AACT, SAA1, and SAA2. Some of these were also detected in our study; however, we found that SAA1 is more associated with non-infectious inflammation patients [25]. ORM1, or Alpha-1-acid glycoprotein 1, is a protein that has an important association with sepsis (0,64) in our study. It's a protein produced by the liver and peripheral tissues in response to systemic inflammation, with immunomodulatory capabilities and the ability to transport substances related to inflammation [26]. Lu et al. observed that ORM1 expression was upregulated in septic patients compared to non-septic patients at the time of discharge [27]. Similar results were observed in a study with 25 septic patients compared to 25 healthy controls [28]. Liang X et al. compared the proteome of a total of 114 patients with sepsis and 62 normal controls in a study. They detected that 81 proteins were involved in different metabolic pathways, such as the acute inflammatory response, platelet degranulation, immune response, and lipid metabolism, among others. Of these proteins, CRP, SAA1, LBP, and A2GL presented highly specific expression levels and could be used as sensitive biomarkers to diagnose patients with sepsis [29]. García-Obregón S et al. analyzed samples from 85 patients with sepsis and 67 healthy donors. After proteomic analysis, they found 6 proteins responsible for inflammation and immune response. Among them, SAA1 was upregulated in septic patients while three other observed proteins (APOE, FHR1 and Hb-β) did not find differences between septic patients and healthy volunteers [30]. In our study, we have observed that APOE is a biomarker that is more associated with NISIRS than with sepsis. Soares AJ et al. examined alterations in the plasma proteome in ten adults affected by sepsis caused by *Acinetobacter baumannii* compared to matched healthy controls. A total of 19 proteins were identified that belong to the inflammation/coagulation pathways and the kallikrein-kinin system. Of these, 10 were upregulated (including C3 and group of SAA proteins) and 9 were decreased (including AHSG) [31], although in our study this protein is more associated with NISIRS.

Despite the fact that in most of the reviewed studies the observed proteins belong to different metabolic pathways, some studies have identified specific proteomes of particular metabolic pathways. Lipopolysaccharides (LPS) are molecules located in the outer membrane of gram-negative bacteria and act as potent activators of the immune system by being recognized by toll-like receptor 4, present in cells such as monocytes, macrophages, neutrophils, and dendritic cells [32]. In our study, we identified two proteins associated with the response to lipopolysaccharide stimulation: PPBP and LBP. Among them, PPBP shows a strong association with sepsis (1,31), while LBP's association is significantly weaker (0,51). In a study conducted by Qia WJ et al., LPS was administered to healthy volunteers, and through proteomic techniques, an upregulation of several inflammatory mediators was observed, such as lipopolysaccharide-binding protein (LBP), CRP, serum amyloid A (SAA), and vWF. Furthermore, analysis at 9 hours after the first LPS injection showed that these three proteins continued to be positively regulated consistently [33]. Kwon OK et al. observed that 19 proteins are secreted by endothelial cells after being stimulated with LPS including C3 and AHSG [34]. Jiao J et al., in a study conducted with rats subjected to cecal ligation and puncture (CLP), found 47 proteins, among which PPBP, Ficolin 1, APOB, LBP, and SERPINA3 were upregulated, while C3 was downregulated in septic rats compared to control rats, correlating closely with the diagnosis of sepsis [35]. Tang et al., using proteomic techniques, were able to isolate 7 proteins, including PPBP. In vitro, these proteins were exposed to cultures with different bacteria (*E. coli* and *Staphylococcus aureus*) and fungi (*Cryptococcus neoformans* and *Candida albicans*), and they observed good antimicrobial activity against the first three pathogens but not against *C. albicans* [36]. In our study, we observed that PPBP is the protein that presents the highest association with sepsis and has been proposed as a biomarker for sepsis [37].

The role of lipoproteins in sepsis is increasingly recognized as a fundamental aspect of the early host response to infection [38]. In a proteomic study on sepsis, Li et al. identified several proteins, including APOE. They observed that the expression of this lipoprotein increases compared to non-septic patients, with an AUC of 0.619 (95% CI: 0.510-0.719) [39]. Additionally, other proteins that act in different metabolic pathways, such as CA1 and SAA1, were also found to be upregulated. Cao et al. collected plasma from patients with community-acquired pneumonia (CAP) aged 50–65 years and 70–85 years, with and without sepsis, as samples for semi-quantitative plasma proteomics. They identified fifty-eight proteins involved in various metabolic pathways. Regarding lipid metabolism, they observed an increase in expression of Apo B100 and Apo E, while Apo C and Apo A were downregulated. Furthermore, in the age group analysis, ApoE levels were higher in younger adults, while ApoA and ApoC were higher in the older age group [40]. In another study involving 27 patients hospitalized for hospital-acquired pneumonia (HAP), the plasma proteome study found that lipid metabolism emerged as the main altered function in these patients compared to healthy volunteers, with High density lipoprotein (HDL) as the central node in the network analysis, interacting with downregulated proteins such as APOA4, APOB, APOC1, APOL1, SAA4, and Paraoxonase 1 (PON1). Validation tests showed reduced plasma levels of total cholesterol, HDL-C, LDL-C, non-HDL cholesterol, apolipoproteins ApoA1 and ApoB100, and PON1 in HAP patients [41]. Kumaraswamy et al. observed that both severe sepsis and NISIRS led to a decrease in blood levels of APOB, although in septic patients, the decrease was significantly greater [42]. In our study, two lipoproteins (APOE and APOH) were found, although we observed that their role is greater in NISIRS than in sepsis, so future studies are necessary to confirm or not the findings.

In sepsis, impaired blood coagulation and thrombus formation are an aspect of the host defense mechanisms [43]. In our study, we observed four proteins associated with coagulation, of which von Willebrand factor (vWF) plays the most significant role in sepsis. In the previously mentioned study by Cao et al., they also found the expression of proteins associated with coagulation pathways. Elevated levels of fibrinogen alpha chain, fibrinogen beta chain, fibrinogen gamma chain, and vWF were found in young adults who developed severe sepsis. Additionally, they found lower levels of antithrombin III (ATIII). Elevated levels of vWF are associated with increased mortality in sepsis [40]. In a study involving septic patients and healthy controls, Kremer Hovinga et al. found significantly higher levels of vWF antigen in septic patients, with no relation to severity or prognosis [44]. Singh et al. confirmed these findings, additionally observing that vWF levels were higher in those who did not survive sepsis [45]. Liu and collaborators utilized two-dimensional electrophoresis (2-DE) and MALDI-TOF-MS mass spectrometry techniques to identify platelet proteins whose expression differs between patients with sepsis and healthy individuals. The study results showed that five platelet proteins had increased expression in patients with sepsis: EFCAB7 (involved in calcium ion binding), actin (part of the cytoskeleton), IL-1b (a cytokine), GPIX (a membrane receptor), and GPIIb (an integrin). These proteins participate in inflammatory response processes and coagulation activation, emphasizing the fundamental role of platelets in the inflammation and coagulation triggered by sepsis [46]. In a study on patients with sepsis due to CAP, several proteins related to the inflammatory response were identified, including SAA1, ORM1, ATM, SERPINA1, SERPINA3, CRP, and LBP (upregulated) and F2, GSN, and APOE (downregulated). Some of these proteins are also key molecules in coagulation and bleeding processes [47].

In sepsis, the inflammatory reaction can damage organs such as the lungs, heart, and kidneys, increasing the risk of mortality. The biological processes that trigger this damage are not yet fully understood, and without specific biomarkers to diagnose or predict the progression of the disease, it can advance to a state of septic shock and even be fatal. To find these biomarkers, various studies have analyzed the proteome in the organs that become damaged during sepsis. Star BS et al. identified nine proteins, including carbonic anhydrase (CA1), which allowed differentiation between patients with sepsis admitted to the emergency department who developed acute kidney injury (AKI) within the first 24 hours and those septic patients who did not present AKI, suggesting that these could act as early biomarkers for the detection of sepsis-AKI [48]. Hashida et al. studied 20 patients with AKI upon admission to the ICU who received continuous renal replacement therapy, 10 of whom had sepsis. Protein was extracted from the adsorbates of the hemofilter and analyzed using proteomic techniques, identifying 197 proteins, of which 9 (including CA1) were found in all studied septic patients compared to non-septic patients [49]. Hinkelbein J et al. conducted a study in rats where they examined the proteome of cardiac tissue after subjecting the animals to a sepsis model induced by CLP. In their findings, they identified 12 proteins in the heart that showed significant alterations. Of these proteins, five (acyl-CoA synthetase 2-like, E1 component of 2-oxoglutarate dehydrogenase, oxoglutarate dehydrogenase, 2-oxoglutarate dehydrogenase complex, and succinate coenzyme A ligase) showed reduced expression 48 hours after sepsis was initiated. The decrease in these proteins is associated with a reduced capacity of the heart to produce energy adequately [50]. In our study, we did not focus on analyzing how protein expression varies specifically in relation to organ dysfunction. However, we consider this aspect crucial and an important starting point for future research. The lack of this analysis in our work indicates that there is significant place to explore how alterations in protein expression may relate to the functioning or dysfunction of different organs. This approach could provide a deeper understanding of the biological mechanisms underlying diseases and open new avenues for diagnosis and treatment.

This study has several limitations. The first and main is that it only allows us to determine the degree of association that these proteins have with sepsis compared to patients with NISIRS, but we do not know the concentrations at which they are present in the patients. We cannot determine whether the expression is upregulated or downregulated compared to other studies.

Second, this is a single-center study. Our results only have internal validity due to the demographic characteristics of the patients. Therefore, it would be necessary to conduct multicenter studies to validate and generalize the findings found in this study.

Third, in septic patients group, samples are collected upon sepsis code activation, and although they are taken early in the course of sepsis, we cannot rule out that all patients present the same stage of evolution at a pathophysiological level at the time of sample collection, which could affect the results.

**CONCLUSION**

There are proteomic patterns associated with sepsis compared to NISIRS with different strength of association. Advances in understanding these protein changes may allow for the identification of new biomarkers or therapeutics targets based in precision medicine in the future.

**Abbreviations**

AUC: Area under the curve

CAP: Community-acquired pneumonia

CLP: Cecal ligation and puncture

CRP: C-reactive protein

FDR: False discovery rate

HAP: Hospital-acquired pneumonia

HDL: High density lipoprotein

ISC: Intra-hospitalalty sepsis code

KNN: K-nearest neighbor

LPS: Lipopolysaccharides

MS: Mass spectrometry

NISIRS: Non-infectious systemic inflammatory response syndrome

PCT: Procalcitonin

qSOFA: Quick SOFA

RFE: Recursive feature elimination

SIRS: Systemic inflammatory response syndrome

SOFA: Sequential Organ Failure Assessment

SVC: Support vector classifier

**Acknowledgements**

To Toni del Pino, Rosa Ras and Pol Herrero from the Proteomics and Metabolomics Area of the Center for Omic Sciences (COS), a Joint between Rovira I Virgili University and Eurecat (Reus, Spain), for their contribution to the proteomics analysis. Samples and data from patients included in this study were provided by Sepsis Bank of the Vall d'Hebron University Hospital Biobank (PT20/00107), integrated in the Spanish National Biobanks Network, and they were processed following standard operating procedures with the appropriate approval of the Ethical and Scientific Committees. The authors kindly appreciate the generous donation of samples and clinical data of the donors of the Sepsis Bank of HUVH Biobank.

**Author contributions**

A.R-S, J.C.R-R.: Conceptualization, Data curation, Formal analysis, Investigation, Methodology, Supervision, Validation, Writing – original draft, Writing – review & editing; V.R.: Conceptualization, Formal analysis, Investigation, Methodology, Supervision, Validation, Writing – original draft , Writing – review & editing; D.S.: Formal analysis, Methodology; L.C-C, L.M., I.B.: Data curation, Formal analysis, Investigation, Validation; N.L., J.J.G., M.D.C.: Validation; N.C.: Formal analysis, Investigation, Methodology, Validation, Writing – original draft; R.F.: : Supervision, Validation, Writing – review & editing.

**Funding**

This study has been funded by Eurecat 2017 Research Projects (Health Forecast 2.0. Omic stratification of patients with sepsis and septic shock).

**Availability of data and materials**

The datasets used and analyzed in the current study are available from the corresponding author upon reasonable request.

## DECLARATIONS

**Ethics approval and consent to participate**

The study was approved by the Clinical Research Ethics Committee of Vall d'Hebron University Hospital [PR (AG) 11-2016, PR (AG) 336-2016, PR (AG) 210/2017], and written informed consent was obtained from all participants.

**Consent for publication**

Consent to publish has been obtained from patients or their relatives.

**Competing interests**

All authors declare no conflicts of interest.

**Author details**

1. Departament de Medicina, Universitat Autònoma de Barcelona, Barcelona, Spain. 2. Intensive Care Department, Vall d'Hebron University Hospital, Vall d'Hebron Barcelona Hospital Campus, Barcelona, Spain. 3. Shock, Organ Dysfunction and Resuscitation (SODIR) Research Group, Vall d'Hebron Research Institute, Barcelona, Spain. 4. Eurecat, Centre Tecnològic de Catalunya, Digital Health Unit, Barcelona, Spain. 5. Department of Clinical Microbiology, Vall d'Hebron University Hospital, Vall d'Hebron Barcelona Hospital Campus, Barcelona, Spain. 6. Eurecat, Centre Tecnològic de Catalunya, Centre for Omic Sciences (COS), Joint Unit URV-EURECAT, Unique Scientific and Technical Infrastructures (ICTS), Reus, Spain. 7. Department of Genetics and Microbiology, Universitat Autònoma de Barcelona, Barcelona, Spain. 8. CIBERINFEC, ISCIII – CIBER de Enfermedades Infecciosas, Instituto de Salud Carlos III, Madrid, Spain. 9. Post-cardiac Surgery Unit. Department of Intensive Care, Vall d'Hebron University Hospital, Vall d'Hebron Barcelona Hospital Campus, Barcelona, Spain.

## FIGURE FOOTNOTES

**Fig. 1** Network of physiological interactions between the different proteins analyzed (strings) (https://string-db.org/).

**Fig. 2** Shap values graphic for logistic regression coefficients. Association strength of each protein in the overall set of patients studied.

**Fig. 3** Coefficient values indicating tendency towards Sepsis (Blue) or NISIRS (Orange)

**REFERENCES**

1. Singer M, Deutschman CS, Seymour CW. The third international consensus definitions for sepsis and septic shock (Sepsis-3). JAMA. 2016;315(8):801-10.
2. Pavon A, Binquet C, Kara F, et al; EPIdemiology of Septic Shock (EPISS) Study Group: Profile of the risk of death after septic shock in the present era: An epidemiologic study. Crit Care Med. 2013;41:2600-9.
3. Yébenes JC, Ruiz-Rodriguez JC, Ferrer R, Clèries M, Bosch A, Lorencio C et al. Epidemiology of sepsis in Catalonia: analysis of incidence and outcomes in a European setting. Ann Intensive Care. 2017 Dec;7(1):19.
4. Vincent JL, Moreno R, Takala J, Willatts S, De Mendonca A, Bruining H, et al. The SOFA (Sepsis- related Organ Failure Assessment) score to describe organ dysfunction/failure. On behalf of the Work- ing Group on Sepsis-Related Problems of the European Society of Intensive Care Medicine. Intensive Care Med. 1996;22(7):707-10.
5. Jiang J, Yang J, Mei J, Jin Y, Lu Y. Head-to-head comparison of qSOFA and SIRS criteria in predicting the mortality of infected patients in the emergency department: a meta-analysis. Scand J Trauma Resusc Emerg Med. 2018 Jul 11;26(1):56.
6. Ginn AN, Halliday CL, Douglas AP, Chen SC. PCR-based tests for the early diagnosis of sepsis. Where do we stand? Curr Opin Infect Dis. 2017 Dec;30(6):565-72.
7. Komiya K, Ishii H, Teramoto S, Takahashi O, Yamamoto H, Oka H, Umeki K, Kadota J. Plasma C-reactive protein levels are associated with mortality in elderly with acute lung injury. J Crit Care. 2012 Oct;27(5):524e1-6.

8. Ruiz-Rodríguez JC, Caballero J, Ruiz-Sanmartín A, Ribas VJ, Pérez M, Bóveda JL, Rello J. Usefulness of procalcitonin clearance as a prognostic biomarker in septic shock. A prospective pilot study. Med Intensiva. 2012 Oct;36(7):475-80.

9. Wirz Y, Meier MA, Bouadma L, Luyt CE, Wolff M, Chastre J, Tubach F, Schroeder S, Nobre V, Annane D, Reinhart K, Damas P, Nijsten M, Shajiei A, deLange DW, Deliberato RO et al. Effect of procalcitonin-guided antibiotic treatment on clinical outcomes in intensive care unit patients with infection and sepsis patients: a patient-level meta-analysis of randomized trials. Crit Care. 2018 Aug 15;22(1):191.

10. Becker KL, Snider R, Nylen ES. Procalcitonin assay in systemic inflammation, infection, and sepsis: clinical utility and limitations. Crit Care Med. 2008 Mar;36(3):941-52.

11. Sager R, Kutz A, Mueller B, Schuetz P. Procalcitonin-guided diagnosis and antibiotic stewardship revisited. BMC Med. 2017 Jan 24;15(1):15.

12. Memar MY, Baghi HB. Presepsin: A promising biomarker for the detection of bacterial infections. Biomed Pharmacother. 2019 Mar;111:649-56.

13. Baldirà J, Ruiz-Rodríguez JC, Wilson DC, Ruiz-Sanmartín A, Cortes A, Chiscano L, Ferrer-Costa R, Comas I, Larrosa N, Fàbrega A, González-López JJ, Ferrer R. Biomarkers and clinical scores to aid the identification of disease severity and intensive care requirement following activation of an in-hospital sepsis code. Ann Intensive Care. 2020 Jan 15;10(1):7.

14. Baldirà J, Ruiz-Rodríguez JC, Ruiz-Sanmartin A, Chiscano L, Cortes A, Sistac DÁ, Ferrer-Costa R, Comas I, Villena Y, Larrosa MN, González-López JJ, Ferrer. Use of Biomarkers to Improve 28-Day Mortality Stratification in Patients with Sepsis and SOFA ≤ 6. R Biomedicines. 2023 Jul 30;11(8):2149.

15. Chiscano-Camón L, Plata-Menchaca E, Ruiz-Rodríguez JC, Ferrer R. [Pathophysiology of septic shock. Med Intensiva (Engl Ed). 2022 Apr:46 Suppl 1:1-13.

16. List EO, Berryman DE, Bower B, Sackmann-Sala L, Gosney E et al. The use of proteomics to study infectious diseases. Infect Disord Drug Targets. 2008;8:31-45.

17. Liu X, Ren H, Peng D. Sepsis biomarkers: an omics perspective. Front Med. 2014 Mar;8(1):58-67.

18. Blangy-Letheule A, Persello A, Rozec B, Waard M, Lauzier B. New approaches to identify sepsis biomarkers: the importance of model and sample source for mass spectrometry. Oxid Med Cell Longev. 2020 Dec 24;2020:6681073.

19. Ruiz-Sanmartín A, Ribas V, Suñol D, Chiscano-Camón L, Palmada C et al. Characterization of a proteomic profile associated with organ dysfunction and mortality of sepsis and septic shock. PLoS One. 2022 Dec 2;17(12):e0278708.

20. Ferrer R, Ruiz-Rodriguez JC, Larrosa N, Llaneras J, Molas E, González-López JJ. Sepsis code implementation at Vall d'Hebron university hospital: rapid diagnostics key to success. ICU Manag Pract. 2017: 17 (4):214-15.

21. Bone RC, Balk RA, Cerra FB, et al. American College of Chest Physicians/Society of Critical Care Medicine Consensus Conference: definitions for sepsis and organ failure and guidelines for the use of innovative therapies in sepsis. Crit Care Med. 1992; 20(6):864-74.

22. Shen Z, Want EJ, Chen W, Keating W, Nussbaumer W, Moore R et al. Sepsis plasma protein profiling with immunodepletion, three-dimensional liquid chromatography tandem mass spectrometry, and spectrum counting. Proteome Res. 2006 Nov;5(11):3154-60.

23. Su L, Zhou R, Liu C, Wen B, Xiao K, Kong W et al. Urinary proteomics analysis for sepsis biomarkers with iTRAQ labeling and two-dimensional liquid chromatography-tandem mass spectrometry. Trauma Acute Care Surg. 2013 Mar;74(3):940-5.

24. Stöve S, Welte T, Wagner TO, Kola A, Klos A, Bautsch W, Köhl J. Circulating complement proteins in patients with sepsis or systemic inflammatory response syndrome. Clin Diagn Lab Immunol. 1996 Mar;3(2):175-83.

25. Chen Q, Liang X, Wu T, Jiang Y, Zang S et al. Integrative analysis of metabolomics and proteomics reveals amino acid metabolism disorder in sepsis. J Transl Med. 2022 Mar 14;20(1): 123.

26. Ceciliani F, Lecchi C. The Immune Functions of α1 Acid Glycoprotein. Curr Protein Pept Sci. 2019;20(6):505-24.

27. Lu J, Li Q, Wu Z, Zhong Z, Ji P, Li H, He C, Feng J, Zhang J. Two gene set variation indexes as potential diagnostic tool for sepsis. Am J Transl Res. 2020 Jun 15;12(6):2749-59.

28. Lu J, Chen R, Ou Y, Jiang Q, Wang L, Liu G, Liu Y, Yang B, Zhou Z, Zuo L, Chen Z. Characterization of immune-related genes and immune infiltration features for early diagnosis, prognosis and recognition of immunosuppression in sepsis. Int Immunopharmacol. 2022 Jun;107:108650.

29. Liang X, Wu T, Chen Q, Jiang J, Jiang Y, Ruan Y, Zhang H et al. Serum proteomics reveals disorder of lipoprotein metabolism in sepsis. Life Sci Alliance. 2021 Aug 24;4(10):e202101091.

30. Garcia-Obregon S, Azkargorta M, Seijas I, Pilar-Orive J, Borrego F, Elortza F et al. Identification of a panel of serum protein markers in early stage of sepsis and its validation in a cohort of patients. J Microbiol Immunol Infect. 2018 Aug;51(4):465-472.

31. Soares AJ, Santos MF, Trugilho MR, Neves-Ferreira AG, Perales J, Domont GB. Differential proteomics of the plasma of individuals with sepsis caused by Acinetobacter baumannii. J Proteomics. 2009 Dec 1;73(2):267-78.

32. Wang X, Quinn PJ. Lipopolysaccharide: Biosynthetic pathway and structure modification. Prog Lipid Res. 2010 Apr;49(2):97-107.

33. Qian WJ, Jacobs JM, Camp DG 2nd, Monroe ME, Moore RJ, Gritsenko MA et al. Comparative proteome analyses of human plasma following in vivo lipopolysaccharide administration using multidimensional separations coupled with tandem mass spectrometry. Proteomics. 2005 Feb;5(2):572-84.

34. Kwon OK, Lee W, Kim SJ, Lee YM, Lee JY, Kim JY et al. In-depth proteomics approach of secretome to identify novel biomarker for sepsis in LPS-stimulated endothelial cells. Electrophoresis. 2015 Dec;36(23):2851-8.

35. Jiao J, Gao M, Zhang H, Wang N, Xiao Z, Liu K et al. Identification of potential biomarkers by serum proteomics analysis in rats with sepsis. Shock. 2014 Jul;42(1):75-81.

36. Tang YQ, Yeaman MR, Selsted ME. Antimicrobial peptides from human platelets. Infect Immun. 2002 Dec;70(12):6524-33.

37. Smith NL, Bromley MJ, Denning DW, Simpson A, Bowyer P. Elevated levels of the neutrophil chemoattractant pro-platelet basic protein in macrophages from individuals with chronic and allergic aspergillosis. J Infect Dis. 2015 Feb 15;211(4):651-60.

38. Harris HW, Gosnell JE, Kumwenda ZL. The lipemia of sepsis: triglyceride-rich lipoproteins as agents of innate immunity. J Endotoxin Res. 2000;6(6):421-30.

39. Li M, Ren R, Yan M, Chen S, Chen C, Yan J. Identification of novel biomarkers for sepsis diagnosis via serum proteomic analysis using iTRAQ-2D-LC-MS/MS. J Clin Lab Anal. 2022 Jan;36(1):e24142.

40. Cao Z, Yende S, Kellum JA, Angus DC, Robinson RA. Proteomics reveals age-related differences in the host immune response to sepsis. J Proteome Res. 2014 Feb 7;13(2):422-32.

41. Sharma NK, Ferreira BL, Tashima AK, Brunialti MKC, Torquato RJS, Bafi A et al. Lipid metabolism impairment in patients with sepsis secondary to hospital acquired pneumonia, a proteomic analysis. Clin Proteomics. 2019 Jul 16;16:29.

42. Kumaraswamy SB, Linder A, Åkesson P, Dahlbäck B. Decreased plasma concentrations of apolipoprotein M in sepsis and systemic inflammatory response syndromes. Crit Care. 2012 Dec 12;16(2):R60.

43. Ono T, Mimuro J, Madoiwa S, Soejima K, Kashiwakura Y, Ishiwata A et al. Severe secondary deficiency of von Willebrand factor-cleaving protease (ADAMTS13) in patients with sepsis-induced disseminated intravascular coagulation: its correlation with development of renal failure. Blood. 2006 Jan 15;107(2):528-34.

44. Kremer Hovinga JA, Zeerleder S, Kessler P, Romani de Wit T, van Mourik JA, Hack CE, ten Cate H, Reitsma PH, Wuillemin WA, Lämmle B. ADAMTS-13, von Willebrand factor and related parameters in severe sepsis and septic shock. J Thromb Haemost. 2007 Nov;5(11):2284-90.

45. Singh K, Kwong AC, Madarati H, Kunasekaran S, Sparring T, Fox-Robichaud AE, Liaw PC, Kretz CA. Characterization of ADAMTS13 and von Willebrand factor levels in septic and non-septic ICU patients. PLoS One. 2021 Feb 19;16(2):e0247017.

46. Liu J, Li J, Den X. Proteomic analysis of diferential protein expression in platelets of septic patients. Mol Biol Rep. 2014 May;41(5):3179-85.

47. Sharma NK, Tashima AK, Brunialti MKC, Ferreira ER, Torquato RJS, Mortara RA et al. Proteomic study revealed cellular assembly and lipid metabolism dysregulation in sepsis secondary to community-acquired pneumonia. Sci Rep. 2017 Nov 15;7(1):15606.

48. Star BS, Boahen CK, van der Slikke EC, Quinten VM, Ter Maaten JC, Henning RH et al. Plasma proteomic characterization of the development of acute kidney injury in early sepsis patients. Sci Rep. 2022 Nov 16;12(1):19705.

49. Hashida T, Nakada TA, Satoh M, Tomita K, Kawaguchi R, Nomura F et al. Proteome analysis of hemofilter adsorbates to identify novel substances of sepsis: a pilot study. J Artif Organs. 2017 Jun;20(2):132-137.

50. Hinkelbein J, Kalenka A, Schubert C, Peterka A, Feldmann RE Jr. Proteome and metabolome alterations in heart and liver indicate compromised energy production during sepsis. Protein Pept Lett. 2010 Jan;17(1):18-31.